\documentclass[a4paper,11pt]{article}
\usepackage{pos}

\usepackage{orcidlink}
\usepackage{slashed}
\usepackage{mathtools}
\usepackage{cleveref}
\usepackage{comment}
\usepackage{graphicx}
\usepackage{caption}
\usepackage{subcaption}

\usepackage{amsmath}
\usepackage{amsfonts}

\PassOptionsToPackage{export}{adjustbox}

\def\threshold{t}

\def\pmtt{\texttt{+\!\!\!\raisebox{-.48ex}-}}

\DeclareMathOperator{\re}{Re}
\DeclareMathOperator{\im}{Im}

\def\fulltwoloopfloopsTria{F^{(2,\rm tria)}}

\usepackage{xspace}

\tikzset{
cut/.style={
    opacity=0.4,
    line width=2.5pt,
    shorten <=-15pt,
    shorten >=-15pt}
}
\usepackage[compat=1.1.0]{tikz-feynman}

\def\ttlf{\ttfamily\fontseries{l}\selectfont}

\usepackage{listings}
\DeclareCaptionLabelFormat{boldlisting}{\textbf{Listing~#2}}
\chardef\MyArticleWithColor=\pdfcolorstackinit page direct{0 g}

\title{Axial triangles in $q\bar{q}\to Z\gamma$ at two loops in QCD directly in four dimensions}
\ShortTitle{Axial triangles in $q\bar{q}\to Z\gamma$ at two loops in QCD}

\author[a,b]{Dario Kermanschah}
\author*[a]{Matilde Vicini}

\affiliation[a]{Institute for Theoretical Physics, ETH Zurich, Wolfgang-Pauli-Strasse 27, 8093 Z\"urich, Switzerland}

\affiliation[b]{Rudolf Peierls Centre for Theoretical Physics, Oxford University, Clarendon Laboratory, Parks Road, Oxford OX1 3PU, UK}

\emailAdd{mvicini@phys.ethz.ch}
\emailAdd{d.kermanschah@gmail.com}

\abstract{
We numerically evaluate the two-loop QCD squared matrix element for in $q\bar{q}\to Z$ and $q\bar{q}\to Z\gamma$ with heavy top and bottom quarks circulating in a triangular fermion loop, by simultaneously subtracting infrared, ultraviolet, and threshold singularities directly in loop momentum space.
This computation serves as an explicit demonstration that axial couplings can be included in the final state within the framework of ref.~\cite{Kermanschah:2025wlo}. 
 By formulating the entire calculation in four spacetime dimensions, with anomaly cancellation realised locally in loop momentum space, we bypass the complications associated with treating $\gamma^5$ in dimensional regularisation.}

\FullConference{17th International Symposium on Radiative Corrections: Applications of Quantum Field Theory to Phenomenology (RADCOR2025)\\
5-10 October 2025\\
Puri, India\\}

\renewcommand{\logo}{\relax}

\begin{document}
\maketitle

\section{Introduction}
In these proceedings, we extend the methods of ref.~\cite{Kermanschah:2025wlo} to include the triangular fermion loop contributions to $q\bar{q}\to Z$ and $q\bar{q}\to Z\gamma$ at two loops in QCD.
These contributions vanish in (di)photon production due to Furry's theorem~\cite{Furry:1937zz}.
When the Z boson couples to the triangular fermion loop, only the axial current contributes, and the two opposite charge flows give the same result~\cite{Sterman:1993hfp}.
Moreover, since the axial couplings within each generation are opposite ($a_u =-a_d$, $a_c =-a_s$, $a_t =-a_b$), the triangle contributions cancel pairwise for equal mass quarks, yielding a non-zero result only when the masses differ.
In particular, they vanish in the two-loop amplitude for $q\bar{q}\to Z\gamma$ computed in massless QCD in ref.~\cite{Gehrmann:2011ab}, where the box-fermion loops were first computed analytically.
In a parallel work, we compute them numerically both for light and heavy quark loops, with the methods of~\cite{Kermanschah:2025wlo}.
\par
Two-loop contributions involving triangular fermion loops were computed for the $Z$ boson decay into quarks~\cite{Kniehl:1989bb,Kuhn:1990kp} at order $O(\alpha_s^2)$ and, related by analytic continuation, for its production~\cite{Dicus:1985wx,Gonsalves:1991qn}.
These calculations were crucial to infer the mass of the top quark experimentally.
We will recompute and validate these contributions and present new results for $q\bar{q}\to Z\gamma$ with massless incoming quarks.
Similar contributions have been calculated in ref.~\cite{Ahmed:2020kme} for $q\bar{q}\to Z H$. 
\par
In our computation, the $u,d,c,s$ are treated as massless. Consequently, the only non-vanishing contribution comes from the heavy $t$ and $b$ quarks in the closed fermion loop.
Contributions from each generation must be considered together to cancel the Adler-Bell-Jackiw anomaly~\cite{Bell:1969ts,Adler:1969gk,Adler:1969er,Collins:1978wz}, otherwise an unphysical ultraviolet (UV) divergence remains and the renormalisability of the theory is spoiled.
After summing all contributing diagrams with $t$ and $b$ quarks in the loop, they are finite in four spacetime dimensions and are neither subject to renormalisation nor sensitive to the choice of factorisation scheme.
We compute them using Monte~Carlo numerical integration in loop momentum space.
The UV divergence is cancelled by locally combining the $t$ and $b$ contributions.
Infrared (IR) divergences are removed using local IR counterterms of \cite{Kermanschah:2025wlo,Anastasiou:2020sdt,Anastasiou:2024xvk} and the threshold singularities are locally subtracted following \cite{Kermanschah:2021wbk,Kermanschah:2024utt,Kermanschah:2025wlo,Vicini:2024ecf}.
\section{The calculation}
\begin{figure}[b]
  \centering
  \begin{subfigure}{.3\textwidth}
    \centering
    \includegraphics[height=2.5cm,page=7]{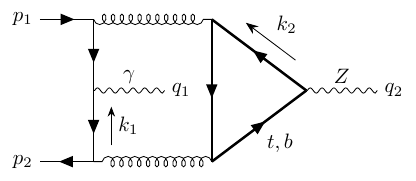}
    \caption{}
    \label{fig:tria_mid}
  \end{subfigure}
  \hfill
  \begin{subfigure}{.3\textwidth}
    \centering
    \includegraphics[height=2.5cm,page=6]{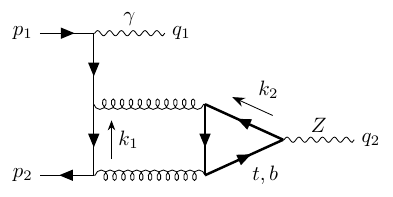}
    \caption{}
    \label{fig:tria_p2}
  \end{subfigure}
  \hfill
  \begin{subfigure}{.3\textwidth}
    \centering
    \includegraphics[height=2.5cm,page=8]{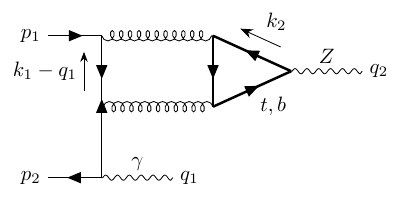}
    \caption{}
    \label{fig:tria_p1}
  \end{subfigure}
\caption{
Representative Feynman diagrams featuring triangular fermion loops with top ($t$) and bottom ($b$) quarks circulating in the loop. 
Diagrams with the opposite charge flow in the fermion loop are not displayed for brevity.
}
  \label{fig:some_diags}
\end{figure}
We consider the processes
\begin{align}
    q(p_1)\, \bar q(p_2) \to \gamma(q_1)\, Z(q_2) \qquad \textrm{and} \qquad   q(p_1)\, \bar q(p_2) \to  Z(q_1)
\end{align}
at two loops in QCD, focusing on the contribution mediated by triangle fermion loops with a quark of flavour $f$ with mass $m_f \geq 0$.
The contributing Feynman diagrams with a single charge flow are shown in fig.~\ref{fig:some_diags}.

Since we consider massless incoming quarks, we need the same local IR counterterm as presented in refs.~\cite{Kermanschah:2025wlo,Anastasiou:2020sdt,Anastasiou:2024xvk}.
In addition, each diagram, taken separately, exhibits local UV divergences. 
However, these UV divergences cancel diagram by diagram when summing over the top and bottom quark contributions in the fermion loop, since the corresponding UV limits are independent of the quark mass. 
As a result, the total contribution shown in fig.~\ref{fig:some_diags} is UV finite ($a_t = - a_b$).
After adding the local IR counterterms, we can therefore perform the numerical integration and Dirac algebra directly in $d = 4$ spacetime dimensions for the sum of the top and bottom quark contributions, without introducing additional UV counterterms. 
The sum of all IR counterterms will integrate to zero due to the Ward identity, as explicitly shown in refs.~\cite{Anastasiou:2020sdt,Anastasiou:2024xvk}.
By integrating finite quantities numerically, we effectively avoid the complications associated with the treatment of $\gamma^5$ in $d$ dimensions that would show up in the corresponding analytic calculation~\cite{Gottlieb:1979ix,Ahmed:2020kme} in dimensional regularisation.
\newcommand{\tba}{{\color{orange}\threshold_1}}
\newcommand{\tcf}{{\color{teal}\threshold_2}}
\newcommand{\tadf}{{\color{magenta}\threshold_3}}
\newcommand{\tbcd}{{\color{teal}\threshold_4^{s}}}
\newcommand{\taeg}{{\color{purple}\threshold_5^{s}}}

\newcommand{\tef}{{\color{purple}\threshold_{s_2}}}
\newcommand{\tbde}{{\color{cyan}\threshold_2^{s_2}}}
\newcommand{\tdg}{{\color{olive}\threshold_3^{s_2}}}
\newcommand{\tacg}{{\color{brown}\threshold_4^{s_2}}}

\newcommand{\tce}{{\color{green}\threshold_1^{s_1}}}
\newcommand{\tade}{{\color{blue}\threshold_2^{s_1}}}
\newcommand{\tbcg}{{\color{pink}\threshold_3^{s_1}}}

\newcommand{\thi}{{\color{red}\threshold^{s_3}}}

\begin{figure}[t]
\centering
\begin{tikzpicture}[scale=2]
\coordinate (a) at (1.75,1);
\coordinate (b) at (2.5,1);
\coordinate (c) at (2.75,0.75);
\coordinate (d) at (2.75,0.25);
\coordinate (e) at (1.75,0);
\draw[cut,orange] (a) -- node[opacity=1,left] {$\tba$} (e);
\draw[cut,teal] (e) -- node[opacity=1,pos=0.24,above] {$\tcf$} (c);
\draw[cut,magenta] (a) -- node[opacity=1,pos=0.24,below] {$\tadf$} (d);
\draw[cut,purple] (d) -- node[opacity=1, left, inner sep=0pt] {$\!\tef$} (c);
\begin{feynman}
\vertex (A) at (0.5,1);
\vertex (B) at (1,1);
\vertex (C) at (2.25,1);
\vertex (D) at (3,1);
\vertex (E) at (3.5,1);
\vertex (F) at (0.5,0);
\vertex (G) at (1,0);
\vertex (H) at (2.25,0);
\vertex (I) at (3,0);
\vertex (J) at (3.5,0);
\vertex (K) at (3,0.5);
\vertex (N) at (3.5,0.5);
\vertex (L) at (1,0.5);
\vertex (M) at (1.5,0.5);
\diagram*{
(A) -- [fermion, momentum=$p_1$] (B),
(K) -- [boson,edge label = {\(Z\)}, momentum'=$q_2$] (N),
(G) -- [fermion, reversed momentum=$p_2$] (F),
(L) -- [boson,edge label = {\(\gamma\)}, momentum'=$q_1$] (M),
(B) -- [gluon] (C),
(G) -- [gluon] (H),
(B) -- [fermion] (L),
(L) -- [fermion, reversed momentum'=$k_1$] (G),
(K) -- [fermion, ultra thick, momentum'=$k_2$] (C),
(H) -- [fermion, ultra thick] (K),
(C) -- [fermion, ultra thick] (H),
};
\end{feynman}
\end{tikzpicture}
\caption{Possible Cutkosky cuts identifying the threshold singularities.}
\label{fig:TriaThresholds}
\end{figure}
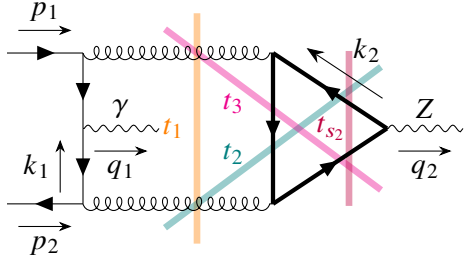
Next, we study the threshold singularities. 
They are analogous to the planar fermion-loop contributions to diphoton production computed in ref.~\cite{Kermanschah:2025wlo}.
The possible threshold singularities are depicted in Fig.~\ref{fig:TriaThresholds} as Cutkosky cuts, and their existence depends on the kinematical parameters.
From these cuts, one can infer the defining equations $t(\vec{k}_1,\vec{k}_2)$ of the associated singular surfaces, after integration over the loop momentum energy components~\cite{Capatti:2019edf}.
For the momentum routing adopted here, one finds, for instance,
\begin{align}
    \tcf(\vec{k}_1,\vec{k}_2):\
    \sqrt{\left|\vec{k}_1-\vec{p}_2\right|^2} + \sqrt{\left|\vec{k}_1+\vec{k}_2+\vec{p}_1-\vec{q}_1\right|^2+m_f^2} +
    \sqrt{\left|\vec{k}_2\right|^2+m_f^2} - p_1^0 - p_2^0+q_1^0= 0\,,
\end{align}
where $m_f$ is the mass of the quark circulating in the fermion loop. 
The existence of real solutions for the spatial loop momenta on these singular surfaces depends on the values of the external kinematics and the internal masses (see, e.g., ref.~\cite{Capatti:2019edf}). 
The kinematical analysis presented in ref.~\cite{Kermanschah:2025wlo} remains applicable in the present case.
For the phase space points in listing~\ref{lst:ps_points}, if $m_f = m_t = 172 \textrm{ GeV}$ we have just $\tba$ active, while if $m_f=m_b = 4.18 \textrm{ GeV}$ we have all the thresholds in fig.~\ref{fig:TriaThresholds} active.

We evaluate the dispersive part of the loop integrals by subtracting carefully-built local threshold counterterms and the absorptive part by integrating back such counterterms~\cite{Kermanschah:2021wbk,Kermanschah:2024utt,Kermanschah:2025wlo}. 

Similar arguments hold for the analogous contributions to $q\bar q \to Z$, which arise from the box–triangle topologies obtained by omitting the photon from figs.~\ref{fig:tria_mid} and \ref{fig:TriaThresholds}.

\section{Numerical results}
The contribution of triangular fermion loops to the helicity-summed two-loop squared matrix element is denoted by
\begin{align}
\label{eq:order}
    \left(\frac{\alpha_s}{4 \pi}\right)^2 C_F T_F \delta_{ij} c_{X} \fulltwoloopfloopsTria, \qquad X\in\{Z,Z\gamma\}
\end{align}
where we factor out the strong coupling $\alpha_s$, the colour factors
\begin{align}
    C_F=\frac{N^2-1}{N}, \qquad T_F=\frac{1}{2}\,,
\end{align}
with $i$ and $j$ the color indices of the incoming quark,
and the boson couplings for $Z$- and $Z\gamma$-production, respectively,
\begin{align}
    c_Z = a_t a_q \qquad c_{Z\gamma} = a_t a_q (eQ_q)^2,
\end{align}
where we have used the relation $a_b = -a_t$, and $eQ_q$ is the electric charge of the incoming quark.
The interference of the axial triangle diagrams with the tree-level amplitude, summed over helicities $\lambda$, then reads
\begin{equation}\label{eq:interf}
    \fulltwoloopfloopsTria \equiv 2\re \sum_{\lambda} M^{(2, \rm tria)}_{\lambda}\left(M_{\lambda}^{(0,A)}\right)^*\,.
\end{equation}
In the two-loop contribution $M^{(2,\rm tria)}$, the $Z$ boson couples only through the axial vertex, whereas at tree level, for a quark of flavour $q$, both axial and vector couplings are present via the structure
\begin{equation}
\label{eq:couplingZ}
   v_q \gamma^\mu - a_q \gamma^\mu \gamma^5 \, .
\end{equation}
Accordingly, the tree amplitude can be decomposed into vector and axial-vector components
\begin{equation}
    v_q \, M^{(0,V)} - a_q \, M^{(0,A)}  \, ,
\end{equation}
where we made explicit the $Z$ boson coupling.
Only the dispersive part of $M^{(2, \rm tria)}$ contributes to $\fulltwoloopfloopsTria$, i.e. 
\begin{equation}
\label{eq:definingInterf}
   \fulltwoloopfloopsTria = 2 \re \sum_{\lambda}
  \textrm{Dispersive}\left[M_{\lambda}^{(2,\rm tria)}\right]\left(M_{\lambda}^{(0,A)}\right)^*\,.
\end{equation}
The contribution proportional to $a_t v_q$ arising from the absorptive part of $M^{(2,\rm tria)}$, i.e.
\begin{equation}
\label{eq:definingInterfAbs}
 -2 \im \sum_{\lambda}
  \textrm{Absorptive}\left[M_{\lambda}^{(2,\rm tria)}\right]\left(M_{\lambda}^{(0,V)}\right)^*\,,
\end{equation}
which could in principle enter the squared matrix element, vanishes.
Indeed, after loop momentum integration and sum over the boson polarisations, there would be a tensor structure with a single Levi-Civita tensor and less than four independent vectors ($p_1,p_2$ for $q\bar{q}\to Z$ and $p_1,p_2,q_1$ for $q\bar{q}\to Z \gamma$). 

First, we consider the numerical evaluation for $q\bar{q}\to Z$ and benchmark it with the analytic expressions of ref.~\cite{Gonsalves:1991qn}. 
The results are presented in table~\ref{tab:znewir}, both with and without numerical stability checks. All implementations yield compatible results. 
We use the default numerical \texttt{double} precision and, when the stability check is activated, we determine the numerical stability of each sampled point using the method described in ref.~\cite{Kermanschah:2024utt}. 
With the stability check enabled, up to one percent of the points are identified as unstable and almost all of them are successfully rescued by re-evaluating the integrand in \texttt{double-double} precision~\cite{double_double,twofloat_dev}. 
For $q\bar{q}\to Z\gamma$, we present the results in table~\ref{tab:table1} for $\fulltwoloopfloopsTria$
evaluated at the three fixed phase-space points of listing~\ref{lst:ps_points}.  

We also verified that the numerical integration of the component from the absorptive part in eq.~\eqref{eq:definingInterfAbs}, proportional to $a_t v_q$, is compatible with zero.

We use the multi-channelling Monte~Carlo and importance sampling developed in ref.~\cite{Kermanschah:2025wlo} and the technical specifications for the code implementation and execution follow closely those described therein.
In the tables below, $\Delta\,[\%]$ denotes the relative precision, $\Delta\,[\sigma]$ the standard deviation in units of the Monte~Carlo error with respect to the reference value, $\mathrm{Exp.}$ the order of magnitude of the result in scientific notation and $N_p$ the number of Monte~Carlo points. 

Table~\ref{tab:znewir} shows the evaluation for $Z$ production, we keep $N_p = 10^9$ fixed and vary $E_\text{CM} = \sqrt{(p_1+p_2)^2}=M_Z$. 
The corresponding reference results for $q\bar q \to Z$ are derived from ref.~\cite{Gonsalves:1991qn}.\footnote{The reference results can be obtained from $\left[G({m_t^2}/{s})-G({m_b^2}/{s})\right]$ from eqs. (2.8), (2.9) and (A.16) of ref.~\cite{Gonsalves:1991qn}.}

Table~\ref{tab:table1} shows the evaluation for $Z\gamma$ production, where we aim to achieve below $0.5\%$ precision, hence $N_p$ can vary.
In listing~\ref{lst:ps_points}, we report the phase space points used for table~\ref{tab:table1}, with $(p_1+p_2)^2 = (1000\,\text{GeV})^2$ and $M_Z = 91.1876\,\text{GeV}$. The masses of the $t$ and $b$ quarks are set to $ m_t = 172\,\text{GeV}$ and $m_b = 4.18\,\text{GeV}$ respectively.

\begin{table}[h]
\centering
\begin{tabular}{rccccc l}
\hline\hline
$E_\text{CM}$ & Exp. & Reference & Numerical result & $\Delta\ [\sigma]$ & $\Delta\ [\%]$ & Stab. \\
\hline\hline
\texttt{100.0} & $\texttt{10}^{\texttt{6}}$ & \ttlf\texttt{ 3.34851} & \ttlf\texttt{ 3.34627} $\pm$ \ttlf\texttt{ 0.00177} & \ttlf\texttt{ 1.260} & \ttlf\texttt{ 0.067} & \texttt{False} \\
\texttt{300.0} & $\texttt{10}^{\texttt{6}}$ & \ttlf\texttt{ 6.55550} & \ttlf\texttt{ 6.55146} $\pm$ \ttlf\texttt{ 0.00696} & \ttlf\texttt{ 0.582} & \ttlf\texttt{ 0.062} & \texttt{False} \\
\texttt{500.0} & $\texttt{10}^{\texttt{7}}$ & \ttlf\texttt{-1.64322} & \ttlf\texttt{-1.64341} $\pm$ \ttlf\texttt{ 0.00125} & \ttlf\texttt{ 0.154} & \ttlf\texttt{ 0.012} & \texttt{False} \\
\texttt{800.0} & $\texttt{10}^{\texttt{7}}$ & \ttlf\texttt{-2.87506} & \ttlf\texttt{-2.87441} $\pm$ \ttlf\texttt{ 0.00261} & \ttlf\texttt{ 0.250} & \ttlf\texttt{ 0.023} & \texttt{False} \\
\texttt{1000.0} & $\texttt{10}^{\texttt{7}}$ & \ttlf\texttt{-3.43074} & \ttlf\texttt{-3.42957} $\pm$ \ttlf\texttt{ 0.00363} & \ttlf\texttt{ 0.323} & \ttlf\texttt{ 0.034} & \texttt{False} \\\hline
\texttt{100.0} & $\texttt{10}^{\texttt{6}}$ & \ttlf\texttt{ 3.34851} & \ttlf\texttt{ 3.34658} $\pm$ \ttlf\texttt{ 0.00177} & \ttlf\texttt{ 1.090} & \ttlf\texttt{ 0.058} & \texttt{True} \\
\texttt{300.0} & $\texttt{10}^{\texttt{6}}$ & \ttlf\texttt{ 6.55550} & \ttlf\texttt{ 6.54795} $\pm$ \ttlf\texttt{ 0.00701} & \ttlf\texttt{ 1.078} & \ttlf\texttt{ 0.115} & \texttt{True} \\
\texttt{500.0} & $\texttt{10}^{\texttt{7}}$ & \ttlf\texttt{-1.64322} & \ttlf\texttt{-1.64284} $\pm$ \ttlf\texttt{ 0.00120} & \ttlf\texttt{ 0.318} & \ttlf\texttt{ 0.023} & \texttt{True} \\
\texttt{800.0} & $\texttt{10}^{\texttt{7}}$ & \ttlf\texttt{-2.87506} & \ttlf\texttt{-2.87275} $\pm$ \ttlf\texttt{ 0.00223} & \ttlf\texttt{ 1.035} & \ttlf\texttt{ 0.080} & \texttt{True} \\
\texttt{1000.0} & $\texttt{10}^{\texttt{7}}$ & \ttlf\texttt{-3.43074} & \ttlf\texttt{-3.42758} $\pm$ \ttlf\texttt{ 0.00298} & \ttlf\texttt{ 1.063} & \ttlf\texttt{ 0.092} & \texttt{True} \\
\hline
\end{tabular}
\caption{Results for $\fulltwoloopfloopsTria_{q\bar q \to Z}$ using as IR counterterms the version in eq.~(3.22, 3.23) of ref.~\cite{Kermanschah:2025wlo}.
 We set $E_\text{CM} = \sqrt{(p_1+p_2)^2}=M_Z$.
 The reference result is obtained from ref.~\cite{Gonsalves:1991qn}.
  These results have been obtained with or without stability check activated as indicated in the last column ({\rm Stab.}). 
  }
\label{tab:znewir}
\end{table}

\begin{table}[h]
\centering
\renewcommand{\arraystretch}{1.35}
\begin{tabular}{lcrcc}
\hline\hline
PSP & Numerical result & Exp. & $N_p [10^8]$ & $\Delta\ [\%]$ \\
\hline\hline
1 & \texttt{-1.4191~$\pmtt$~0.0067} & \ttlf$\texttt{10}^{\texttt{+3}}$ & 1.65& \texttt{0.473} \\
2 & \texttt{-1.1999~$\pmtt$~0.0055} & \ttlf$\texttt{10}^{\texttt{+3}}$ & 1.08 & \texttt{0.459} \\
3 & \texttt{-2.8127~$\pmtt$~0.0134} & \ttlf$\texttt{10}^{\texttt{+3}}$ & 2.80 & \texttt{0.475} \\
\hline\hline
\end{tabular}
\caption{Results for $\fulltwoloopfloopsTria_{q\bar q \to Z\gamma}$.
  These results have been obtained with stability check activated. 
  The label of the phase space points~(PSP) refers to the kinematic configurations in listing~\ref{lst:ps_points}.
  }
  
\label{tab:table1}
\end{table}

\newpage
\begin{lstlisting}[language=Python,frame=single,caption={Phase-space points (PSP) for $q(p_1) \bar{q}(p_2) \to \gamma(q_1) Z(q_2)$ with $\sqrt{(p_1+p_2)^2}=1000\,\text{GeV}$.},commentstyle=\color{teal},captionpos=b,label={lst:ps_points}]
# PSP 1
0.5000000000000000E+03  0.0000000000000000E+00  0.0000000000000000E+00  0.5000000000000000E+03
0.5000000000000000E+03  0.0000000000000000E+00  0.0000000000000000E+00 -0.5000000000000000E+03
0.4958424108776996E+03  0.1100019292470276E+03  0.4411319421848414E+03 -0.1978936117492713E+03
0.5041575891223005E+03 -0.1100019292470276E+03 -0.4411319421848414E+03  0.1978936117492712E+03
# PSP 2
0.5000000000000000E+03  0.0000000000000000E+00  0.0000000000000000E+00  0.5000000000000000E+03
0.5000000000000000E+03  0.0000000000000000E+00  0.0000000000000000E+00 -0.5000000000000000E+03
0.4958424108776997E+03 -0.2795713281286689E+03 -0.3875414700574890E+03  0.1323298072962480E+03
0.5041575891223005E+03  0.2795713281286690E+03  0.3875414700574891E+03 -0.1323298072962480E+03
# PSP 3
0.5000000000000000E+03  0.0000000000000000E+00  0.0000000000000000E+00  0.5000000000000000E+03
0.5000000000000000E+03  0.0000000000000000E+00  0.0000000000000000E+00 -0.5000000000000000E+03
0.4958424108776987E+03 -0.2386202223474536E+03  0.2781398312686913E+03 -0.3340034732958395E+03
0.5041575891223014E+03  0.2386202223474535E+03 -0.2781398312686913E+03  0.3340034732958395E+03
\end{lstlisting}

\section{Conclusion}
In these proceedings we have extended the numerical framework introduced in ref.~\cite{Kermanschah:2025wlo} to include squared matrix element at two loops in QCD involving triangular fermion loops in the processes
$q\bar q \to Z$ and $q\bar q \to Z\gamma$.
For equal quark masses, the axial triangle contributions cancel within each generation. 
Consequently, the only non-vanishing contributions arise from the top and bottom quarks in the quark triangle loop, since they have a non-negligible mass difference.

The integration is performed entirely numerically, using Monte~Carlo methods in loop momentum space.
Ultraviolet divergences cancel locally when summing the top and bottom quark contributions, while infrared singularities are removed by local counterterms.
Threshold singularities are handled by local subtraction. 
By formulating the entire calculation directly in four spacetime dimensions, exploiting local anomaly cancellations, we could effectively avoid the complications associated with the treatment of $\gamma^5$ in dimensional regularisation for these axial triangle contributions.

\acknowledgments
We thank the organisers of RADCOR 2025 for their cordial hospitality.

We gratefully acknowledge the use of the Euler cluster at ETH Zurich for carrying out the numerical computations.
This work was supported by the Swiss National Science Foundation through its Postdoc.Mobility funding scheme (grant number 230593) and project funding scheme (grant number 10001706).
\newpage
\bibliographystyle{JHEP}
\bibliography{biblio.bib}

\end{document}